\documentclass[twocolumn,twoside,slac_two]{revtex4}
\usepackage{graphicx}
\usepackage{fancyhdr}
\begin{document}

\title{High-Energy Neutrinos in Light of Fermi-LAT}

%

\author{Markus Ahlers}
\affiliation{WIPAC \& Department of Physics, University of Wisconsin--Madison, Madison, WI 53706, USA}

\begin{abstract}
The production of high-energy astrophysical neutrinos is tightly linked to the emission of hadronic $\gamma$-rays. I will discuss the recent observation of TeV to PeV neutrinos by the IceCube Cherenkov telescope in the context of $\gamma$-ray astronomy. The corresponding energy range of hadronic $\gamma$-rays is not directly accessible by extragalactic $\gamma$-ray astronomy due to interactions with cosmic radiation backgrounds. Nevertheless, the isotropic sub-TeV $\gamma$-ray background observed by the {\it Fermi} Large Area Telescope (LAT) contains indirect information from secondary emission produced in electromagnetic cascades and constrains hadronic emission scenarios. On the other hand, observation of PeV $\gamma$-rays would provide a smoking-gun signal for Galactic emission. In general, the cross-correlation of neutrino emission with (extended) Galactic and extragalactic $\gamma$-ray sources will serve as the most sensitive probe for a future identification of neutrino sources. 
\end{abstract}

\maketitle

\thispagestyle{fancy}


\section{Introduction}

The recent observation of a flux of high-energy astrophysical neutrinos~\cite{Aartsen:2013bka,Aartsen:2013jdh,Aartsen:2014gkd,Aartsen:2014muf} has added an important new pillar to multi-messenger astronomy. Neutrinos are tracers of hadronic interactions of cosmic rays (CRs) via the production and decay of charged mesons. Unlike the observation of $\gamma$-rays, which can also be produced by leptonic emission, {\it i.e.}~synchrotron emission, bremsstrahlung or inverse-Compton scattering of high-energy electrons, the detection of neutrinos is direct evidence of the presence of high-energy CRs. Due to their weak interaction with matter neutrinos at all energies can arrive from very distant sources and probe the Universe as far as the Hubble horizon. In contrast, $\gamma$-rays at energies beyond a few TeV scatter strongly in cosmic radiation backgrounds and initiate electromagnetic cascades shifting the $\gamma$-ray emission into the sub-TeV region. Cosmic rays are deflected via Galactic and extragalactic magnetic fields and can only correlate with their sources at energies approaching the {\it Greisen-Zatspin-Kuz'min} (GZK) cutoff \cite{Greisen:1966jv,Zatsepin:1966jv}, $E_{\rm GZK} \simeq 50$~EeV. Thus, astronomical observations of non-thermal point sources emitting in the energy band between 10~TeV and 10~EeV are only possible via astrophysical neutrinos.

\begin{figure*}[t]
\centering
\includegraphics[clip=false,viewport=0 10 650 390,width=0.8\linewidth]{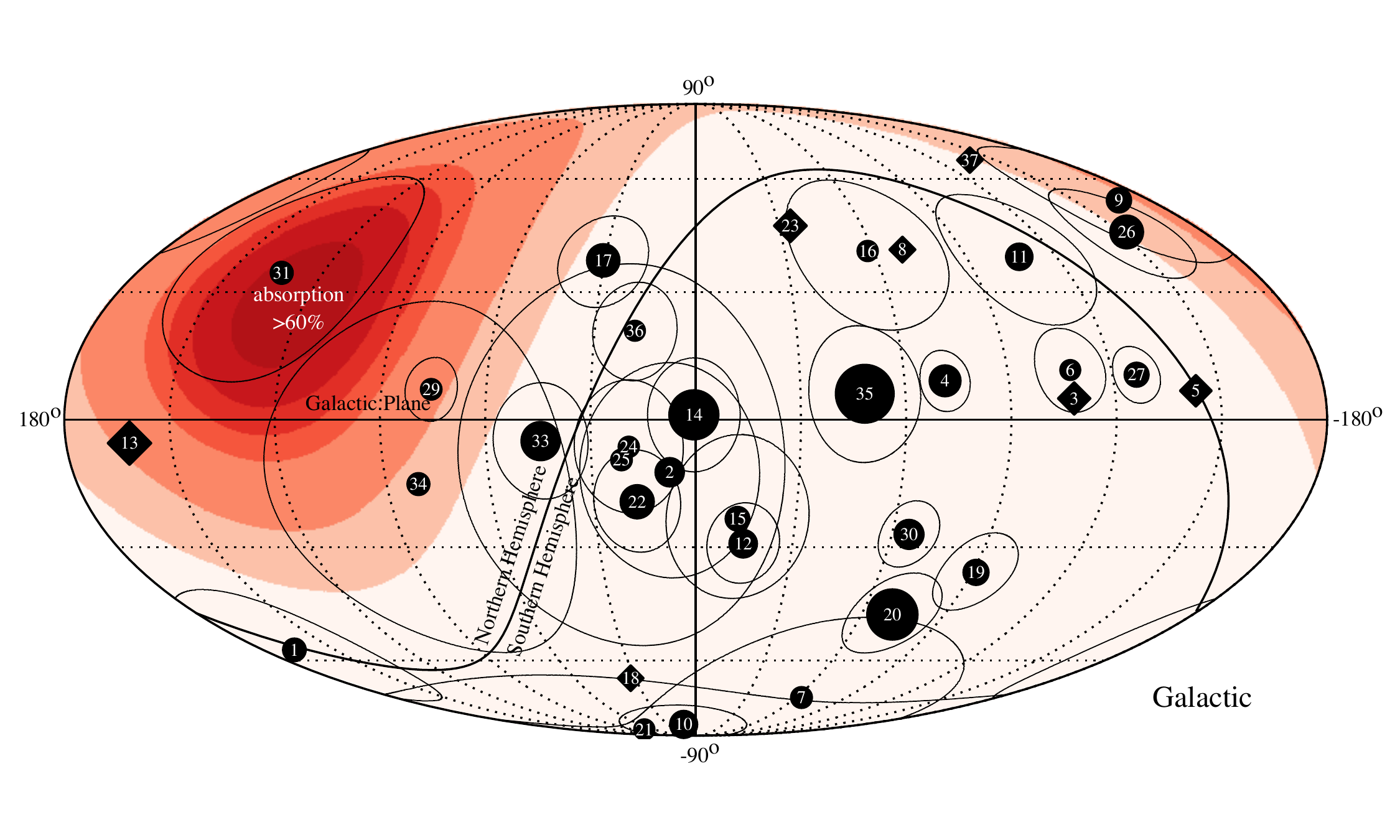}
\caption{The arrival directions of IceCube events from Ref.~\cite{Aartsen:2014gkd}. The events are classified as tracks (diamonds) and cascades (filled circles). The relative detected energy of the events is indicated by the area of the symbols. The thin lines around the arrival direction of the cascade events indicate the systematic uncertainty of the reconstruction. Two likely background events (\#28 and \#35) from the 37 events are omitted from the plot. The red shaded region shows the minimal ($E_\nu=30$~TeV) absorption of the neutrino flux due to scattering in the Earth in 10\% steps.} \label{fig1}
\end{figure*}

On the other hand, the weak interaction of neutrinos with matter is a challenge for their detection requiring enormously large instruments. One possibility consist of the detection of Cherenkov light emitted by high-energy secondary charged particles produced in neutrino interactions in optically transparent media. This is the concept of the IceCube detector which consists of a cubic kilometer of deep glacial ice at the geographic South Pole that is instrumented by an array of digital optical modules (DOMs). The small number of signal events have to compete against large backgrounds from CR activity in the atmosphere producing high-energy muons and atmospheric neutrinos. 

Only recently, the IceCube Collaboration was able to identify a flux of high-energy astrophysical neutrinos~\cite{Aartsen:2013bka,Aartsen:2013jdh,Aartsen:2014gkd,Aartsen:2014muf}. The flux of neutrinos observed in the so-called {\it high-energy starting event} (HESE) analysis consists of 37 events with deposited energies between 30~TeV and 2~PeV observed within a period of three years~\cite{Aartsen:2014gkd}. To extract an astrophysical signal the analysis identifies events with bright Cherenkov light emission of secondary charged particles that passed a virtual outer veto layer of DOMs~\cite{Schonert:2008is}. This does not only veto most of the atmospheric muons, but also a large portion of atmospheric neutrino in the Southern Hemisphere which are vetoed by co-produced shower muons~\cite{Gaisser:2014bja}. The topologies of the HESE events are classified in terms of {\it tracks} and {\it cascades}, depending on whether the neutrino interaction produced a muon track inside the detector or just a nearly spherical emission pattern at its interaction vertex, respectively. The expected number of background events are about 15 atmospheric muons and neutrinos. The total significance of the flux is at $5.7\sigma$~\cite{Aartsen:2014gkd}. 

Figure~\ref{fig1} shows the distribution of the HESE events in Galactic coordinates. The different event topologies of tracks and cascades are shown as diamonds and filled circles, respectively. The area of the symbols indicate the relative increase of deposited energy. The most energetic events consist of three PeV cascades (\#14, \#20 \& \#35). Due to the spherical emission of the cascades the uncertainty in the reconstruction of their initial neutrino arrival direction is typically larger than $10^\circ$ and is indicated as thin circles in the sky map. For tracks the reconstruction has a resolution of better than $1^\circ$. The red shaded area shows $10\%$ steps of the minimal Earth absorption of neutrinos in the sample assuming 30\,TeV as their minimum energy. Accounting for the zenith angle dependence of signal and background the emission is consistent with an isotropic and equal-flavor flux at a level of
\begin{equation}\label{eq:ICflux}
E_\nu^2J^{\rm IC}_{\nu_\alpha} \simeq (0.95\pm0.3)\times10^{-8}{\rm GeV}{\rm s}^{-1}{\rm cm}^{2}{\rm sr}^{-1}\,,
\end{equation}
per neutrino flavor assuming an $E^{-2}$ power-law emission. Track events can only be produced by charged current interactions of muon neutrinos and hence the track-to-cascade ratio contains information of the flavor composition~\cite{Anchordoqui:2004eb,Mena:2014sja,Bhattacharya:2011qu,Barger:2014iua,Winter:2014pya}. A recent analysis of IceCube shows that the observation is consistent with an equal flavor composition expected from astrophysical sources~\cite{Aartsen:2015ivb}. The best-fit spectral index of the HESE analysis is at $2.3$ with an total uncertainty of $\pm0.3$~\cite{Aartsen:2014gkd}. Note, that a recent IceCube analysis extending the veto idea to neutrinos at (1-10)~TeV favors a softer spectrum of $2.46\pm0.12$~\cite{Aartsen:2014muf}.

Various astrophysical scenarios have been suggested that might be (partially) responsible for the observed flux of neutrinos. The absence of significant signs of anisotropy in the data is consistent with an extragalactic population of sources. Source candidates include galaxies with intense star formation~\cite{Loeb:2006tw,Murase:2013rfa,He:2013cqa,Anchordoqui:2014yva,Chang:2014hua,Senno:2015tra}, cores of active galactic nuclei (AGN)~\cite{Stecker:1991vm,Stecker:2013fxa,Kalashev:2014vya}, low-luminosity AGN~\cite{Bai:2014kba,Kimura:2014jba}, blazars~\cite{Tavecchio:2014eia,Padovani:2014bha,Dermer:2014vaa}, low-power GRBs~\cite{Waxman:1997ti,Murase:2013ffa,Ando:2005xi}, cannonball GRBs~\cite{Dado:2014mea}, intergalactic shocks~\cite{Kashiyama:2014rza}, and active galaxies embedded in structured regions~\cite{Berezinsky:1996wx,Murase:2008yt,Murase:2013rfa}. Galactic contributions are in general identifiable by anisotropies in the arrival direction of neutrinos. The data shows no evidence for this, but this might be hidden by the limited event statistics and angular resolution of cascades. Possible contributions to super-TeV neutrinos are the diffuse neutrino emission of galactic CRs~\cite{Ahlers:2013xia,Joshi:2013aua,Kachelriess:2014oma}, the joint emission of galactic PeV sources~\cite{Fox:2013oza,Gonzalez-Garcia:2013iha} or microquasars~\cite{Anchordoqui:2014rca}, and extended galactic structures like the {\it Fermi Bubbles}~\cite{Razzaque:2013uoa,Ahlers:2013xia,Lunardini:2013gva} or the galactic halo~\cite{Taylor:2014hya}. A possible association with the sub-TeV diffuse galactic $\gamma$-ray emission~\cite{Neronov:2013lza} and constraints from the non-observation from diffuse galactic PeV $\gamma$-rays~\cite{Gupta:2013xfa,Ahlers:2013xia}, have also been investigated. More exotic scenarios have suggested a contribution of neutrino emission from decaying heavy dark matter~\cite{Feldstein:2013kka,Esmaili:2013gha,Bai:2013nga,Cherry:2014xra}.

Constraining the origin of the IceCube observation by neutrino data itself is challenging due to low event statistics, large backgrounds and systematic effects. Progress can be made by the fact the neutrino emission is intimately related to the production of hadronic $\gamma$-rays. Observation of $\gamma$-ray astronomy can hence help to constrain or identify the neutrino emission. In particular, the wealth of data coming from the {\it Fermi} telescope which allows for a cross-correlation with neutrino events in IceCube's field of view can help to identify possible sources, as we will discuss in the following.  

\section{Pinpointing Neutrino Sources}

As mentioned in the introduction the neutrino observation is consistent with an isotropic flux. This would naturally arise from a superposition of faint point-sources of an extra-galactic source population. For simplicity, let's consider a distribution of continuously emitting sources with the same emission rate $Q_\nu(E) \propto E^{-\gamma}$ and red-shift dependent density $\mathcal{H}(z)$. The individual point-source spectrum $J$ (in units of ${\rm GeV}^{-1} {\rm s}^{-1} {\rm cm}^{-2}$) at red-shift $z$ is then given as
\begin{equation}
J(z,E) = \frac{(1+z)^2Q_\nu((1+z)E)}{4\pi d^2_L(z)}\,,
\end{equation}
for a luminosity distance $d_L(z)=(1+z)\int{\rm d}z'/H(z')$ defined by the red-shift Hubble expansion rate $H(z)$. In the following we assume a flat universe dominated by vacuum energy with $\Omega_{\Lambda} \simeq 0.7$ and cold dark matter with $\Omega_{\rm m} \simeq 0.3$~\cite{Agashe:2014kda}. The Hubble parameter at earlier times is then given by its value today of $H_0\simeq70$ km s${}^{-1}$ Mpc${}^{-1}$ and the relation $H^2 (z) = H^2_0\,( \Omega_{\Lambda}+\Omega_{\rm m} (1 + z)^3)$. On the other hand, the average diffuse flux of neutrinos originating in multiple cosmic sources is simply given by
\begin{equation}\label{eq:Jtot}
J_{\rm tot}(E_\nu) = \frac{1}{4\pi}\int_0^\infty{{\rm d}z}\frac{{\rm d}\mathcal{V}}{{\rm d}z}\mathcal{H}(z)J_\nu(z,E)\,,
\end{equation}
where $\mathcal{V}(z)=(4\pi/3)d^3_c(z)$ is the co-moving volume with co-moving distance $d_c(z) = d_L(z)/(1+z)$. This quantity is normalized by the diffuse flux of Eq.~(\ref{eq:ICflux}). The contribution of an (average) source at co-moving distance $r$ can then be expressed via the local density $\mathcal{H}_0 = \mathcal{H}(0)$ and an evolution factor
\begin{equation}\label{xi}
\xi_z(E) = \int_0^\infty{\rm d}z\frac{(1+z)^{-\gamma}}{\sqrt{\Omega_\Lambda+(1+z)^3\Omega_{\rm m}}}\frac{\mathcal{H}(z)}{\mathcal{H}(0)}\,.
\end{equation}
Based on the diffuse flux (\ref{eq:ICflux}) we can then estimate the contribution of individual point sources. For a continuously emitting source at a distance $d = d_1 10$~Mpc the mean neutrino flux is given as
\begin{equation}\label{eq:JPS}
E_\nu^2J_\nu 
\simeq \frac{(0.9\pm0.3)\times10^{-12}}{\xi_{z, 2.4}\mathcal{H}_{0, -5}d_1^{2}} \frac{\rm TeV}{{\rm cm}^{2}\,{\rm s}}\,,
\end{equation}
where $\mathcal{H}_0 = \mathcal{H}_{0, -5} 10^{-5} {\rm Mpc}^{-3}$ is the local source density.
An analogous argument can be made for transient sources~\cite{Ahlers:2014ioa}. In this case the time-integrated neutrino flux $F$ (in units of ${\rm GeV}^{-1} {\rm cm}^{-2}$) from an individual transient can be expressed as 
\begin{equation}\label{eq:FPS}
E_\nu^2F_\nu 
\simeq \frac{0.3\pm0.1}{\xi_{z, 2.4}{\dot{\mathcal{H}}}_{0, -6}d_1^{2}} \frac{\rm GeV}{{\rm cm}^{2}}\,,
\end{equation}
where ${\dot{\mathcal{H}}}_0 = {\dot{\mathcal{H}}}_{0, -6}10^{-6} {\rm Mpc}^{-3}{\rm yr}^{-1}$ is the local flaring/burst density rate.

\begin{figure}[t]
\centering
\includegraphics[width=\linewidth]{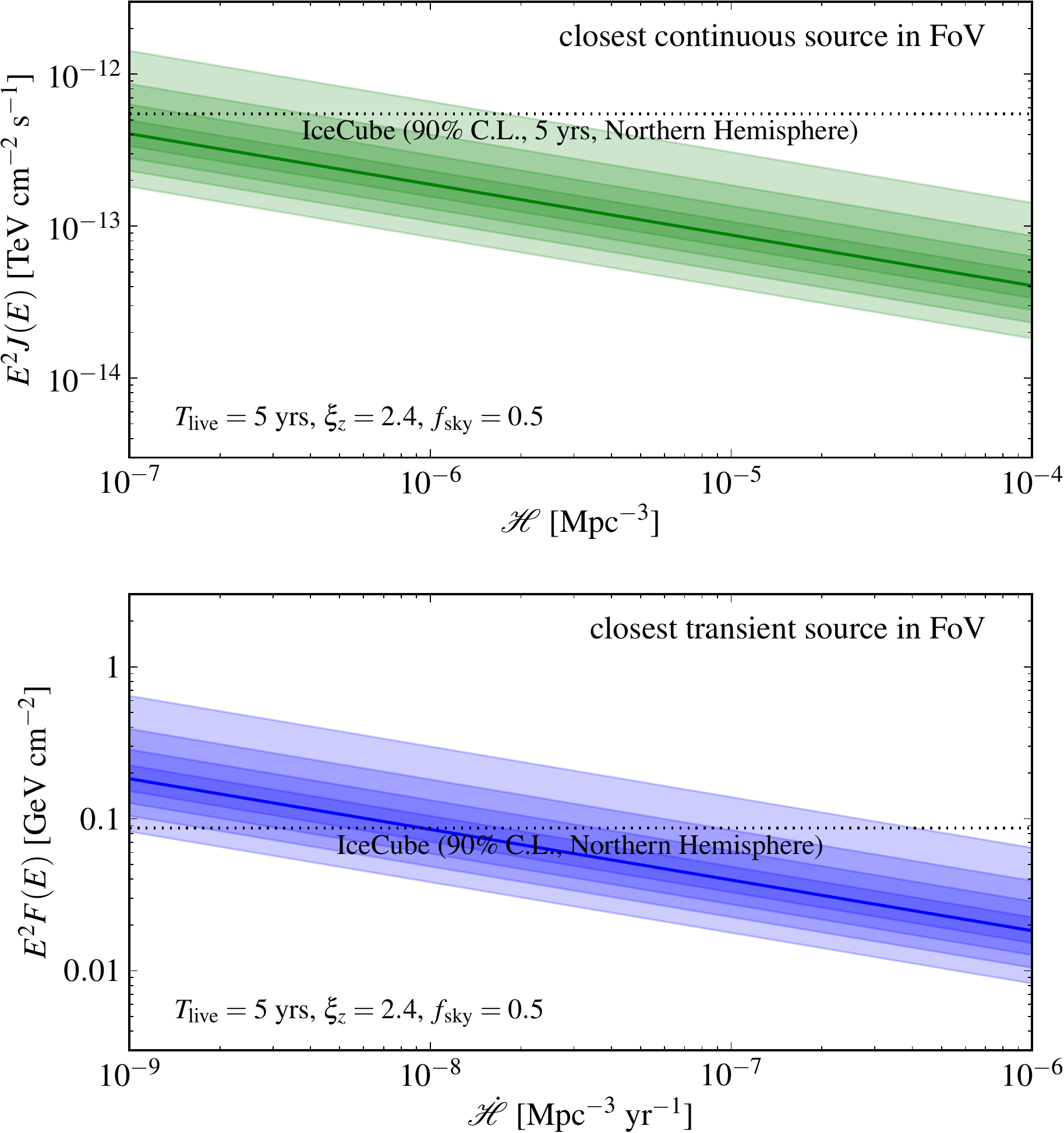}
\caption[]{Expected emission of the closest neutrino source in terms of the average source density. The shaded regions show the 10\% percentiles around the mean (solid line) expected from a random distribution of sources (from Ref.~\cite{Ahlers:2014ioa}).} \label{fig2}
\end{figure}

In Eqs.~(\ref{eq:JPS}) and (\ref{eq:FPS}) the distance $d$ and density $\mathcal{H}$ are kept as independent parameters. However, the first identified neutrino point-source will be the brightest one in the field of view (FoV), {\it i.e.}~the closest one for equal-luminosity sources. The position of the closest source of an ensemble follows a statistical distribution~\cite{Ahlers:2014ioa}. Figure~\ref{fig2} shows the expected flux range of the closest continuous (top) or transient (bottom) neutrino source assuming a homogeneous local distribution with density $\mathcal{H}_0$ or density rate $\dot\mathcal{H}_0$, respectively. The different shaded bands indicate the 10\% percentiles around the mean (solid line). The calculation assumes a source distribution following that of star-formation rate, $\xi_z\simeq 2.4$, using the estimates of Refs.~\cite{Hopkins:2006bw,Yuksel:2008cu}. The plots in Fig.~\ref{fig2} also indicates the point-source sensitivity of IceCube in the Northern Hemisphere after 5 years of observation. IceCube is presently only sensitive to sparse sources with densities of $\mathcal{H}_0\lesssim10^{-7}\,{\rm Mpc}^{-3}$ like flat-spectrum radio quasars or very rare $\dot\mathcal{H}_0\lesssim10^{-8}\,{\rm Mpc}^{-3}\,{\rm yr}^{-1}$ transient source classes like gamma-ray bursts.

Significant progress can be made by cross-correlating neutrino events with source catalogues~\cite{Aartsen:2014cva,Aartsen:2013uuv}. In particular, {\it Fermi} observations of extra-galactic $\gamma$-ray sources with an un-biased FoV provide an excellent catalogue for stacking searches, {\it e.g.}~blazar sources~\cite{Gluesenkamp2014}.  In particular, the large background of atmospheric events can be significantly reduced by searching for neutrino events in coincidence with the position and time of transient sources~\cite{Aartsen:2013uuv}. For instance, IceCube has been looking for neutrino emission in coincidence with gamma-ray bursts (GRBs). The present limit on the combined (``stacked'') emission from GRBs reported via the GRB Coordinates Network~\cite{GRBnetwork} and the {\it Fermi} GBM catalogs over a period of five years places an upper limit on their diffuse muon-neutrino flux which is about 1\% of the observed diffuse emission (\ref{eq:ICflux}), constraining the GRB origin of the emission~\cite{Aartsen:2014aqy}.

\section{Diffuse Neutrinos}

The overall energy density of the observed neutrino flux is close to a theoretical limit for neutrino production in the sources of ultra-high energy (UHE) CRs~\cite{Waxman:1998yy}. This might just be a coincidence, but it can also indicate a multi-messenger relation. The neutrino and CR nucleon ($N$) emission rates $Q$ (in units of ${
\rm GeV}^{-1} {\rm s}^{-1}$) are related via
\begin{equation}\label{eq:QCR}
\frac{1}{3}\sum_{\alpha}E^2_\nu Q_{\nu_\alpha}(E_\nu)  \simeq \frac{1}{4}\frac{f_\pi K_\pi}{1+K_\pi}E^2_NQ_N(E_N)\,
\end{equation}
where $f_\pi<1$ is the pion production efficiency, $K_\pi$ the ratio of charged to neutral pions and $E_\nu\simeq 0.05E_N$. The emission rate density of UHE CRs depend on spectrum and composition. For an $E^{-2}$ flux of protons it can be estimated as $E^2_pQ_p(E_p) \simeq (1-2)\times10^{44}\,{\rm erg}\,{\rm Mpc}^{-3}\,{\rm yr}^{-1}$~\cite{Ahlers:2012rz}. Hence, using Eq.~(\ref{eq:Jtot}) the diffuse neutrino flux can be estimated as
\begin{equation}
E_\nu^2J_\nu(E_\nu) \simeq \frac{\xi_zf_\pi K_\pi}{1+K_\pi}(2-4)\times10^{-8}\,{\rm GeV}\,{\rm cm}^{-2}\,{\rm s}^{-1}\,{\rm sr}\,,
\end{equation}
were $\xi_z$ is again given by Eq.~(\ref{xi}). Since $f_\pi<1$ this provides a theoretical upper limit on neutrino production, the {\it Waxman-Bahcall} (WB) bound~\cite{Waxman:1998yy}. 

Neutrino fluxes close to this limit would require very efficient CR production with optical thickness $\tau_{p\gamma/pp} \gg 1$, such that $f_\pi\simeq 1$, {\it i.e.}~CR reservoirs~\cite{Katz:2013ooa} such as starburst galaxies~\cite{Loeb:2006tw,Tamborra:2014xia} or clusters of galaxies~\cite{Berezinsky:1996wx,Murase:2008yt,Zandanel:2014pva}. Interestingly, the energy density of Galactic CRs require a similar energy density. Assuming that 1\% of the kinetic energy of $10^{51}$~erg of a supernova (SN) explosion is converted to CRs and assuming normal galaxies with densities $\mathcal{H}_0\simeq10^{-3}{\rm Mpc}^{-3}$ and a SN rate of $10^{-2}~{\rm yr}^{-1}$ we arrive at $E^2_pQ_p(E_p)\simeq10^{44}\,{\rm erg}\,{\rm Mpc}^{-3}\,{\rm yr}^{-1}$. This coincidence together with the saturation of the WB bound has let to speculations that Galactic and extragalactic CRs might be produced in the same transient sources~\cite{Katz:2013ooa}. 

Hadronic interactions of CRs will not only produce neutrinos, but also hadronic $\gamma$-rays. The production rates are related by
\begin{equation}\label{eq:Qgamma}
\frac{1}{3}\sum_{\alpha}E^2_\nu Q_{\nu_\alpha}(E_\nu) \simeq \frac{K_\pi}{4}E^2_\gamma Q_\gamma(E_\gamma)\,.
\end{equation}
Note, that this relation does not depend on the pion production efficiency, but only on the relative charged-to-neutral pion rate $K_\pi$. However, the production rate described by Eq.~(\ref{eq:Qgamma}) is not necessarily the emission rate of the sources. For instance, in hadronic sources that efficiently produce neutrinos via $p\gamma$ interactions the target photon field can also efficiently reduce the hadronic $\gamma$-rays via pair production. Inverse-Compton scattering and synchrotron emission in magnetic fields will then shift the emitted $\gamma$-ray spectrum to lower energies. This is a calorimetric process that will conserve the total energy of hadronic $\gamma$-rays.

On the other hand, optically thin sources where the hadronic production is dominated by CR-gas interactions ($pp$ sources) are expected to release the hadronic $\gamma$-rays described by Eq.~(\ref{eq:Qgamma}). For this production mechanism the pion production efficiency is only weakly depend on the initial CR energy. The emitted neutrino and $\gamma$-ray spectra essentially follow the initial power-law spectrum of CRs, {\it cf.}~Eq.~(\ref{eq:QCR}). Nevertheless, the high-energy $\gamma$-rays of extragalactic sources will interact with cosmic radiation backgrounds, in particular the cosmic microwave background. Here again, the pair production and subsequent inverse-Compton scattering of the high energy electrons will lead to electromagnetic cascades. As a result, the initial energy density of hadronic $\gamma$-ray will be shifted into the sub-TeV $\gamma$-ray band, where they supplement the direct emission of the source. The observed $\gamma$-ray background in this energy region provides hence a general upper limit on the diffuse hadronic emission~\cite{Berezinsky:1974kz}, which also applies to the production of cosmogenic neutrinos produced via the GZK interaction~\cite{Berezinsky:1969qj,Berezinsky:2010xa,Ahlers:2010fw,Decerprit:2011qe}.

\begin{figure}[t]
\centering
\includegraphics[width=\linewidth]{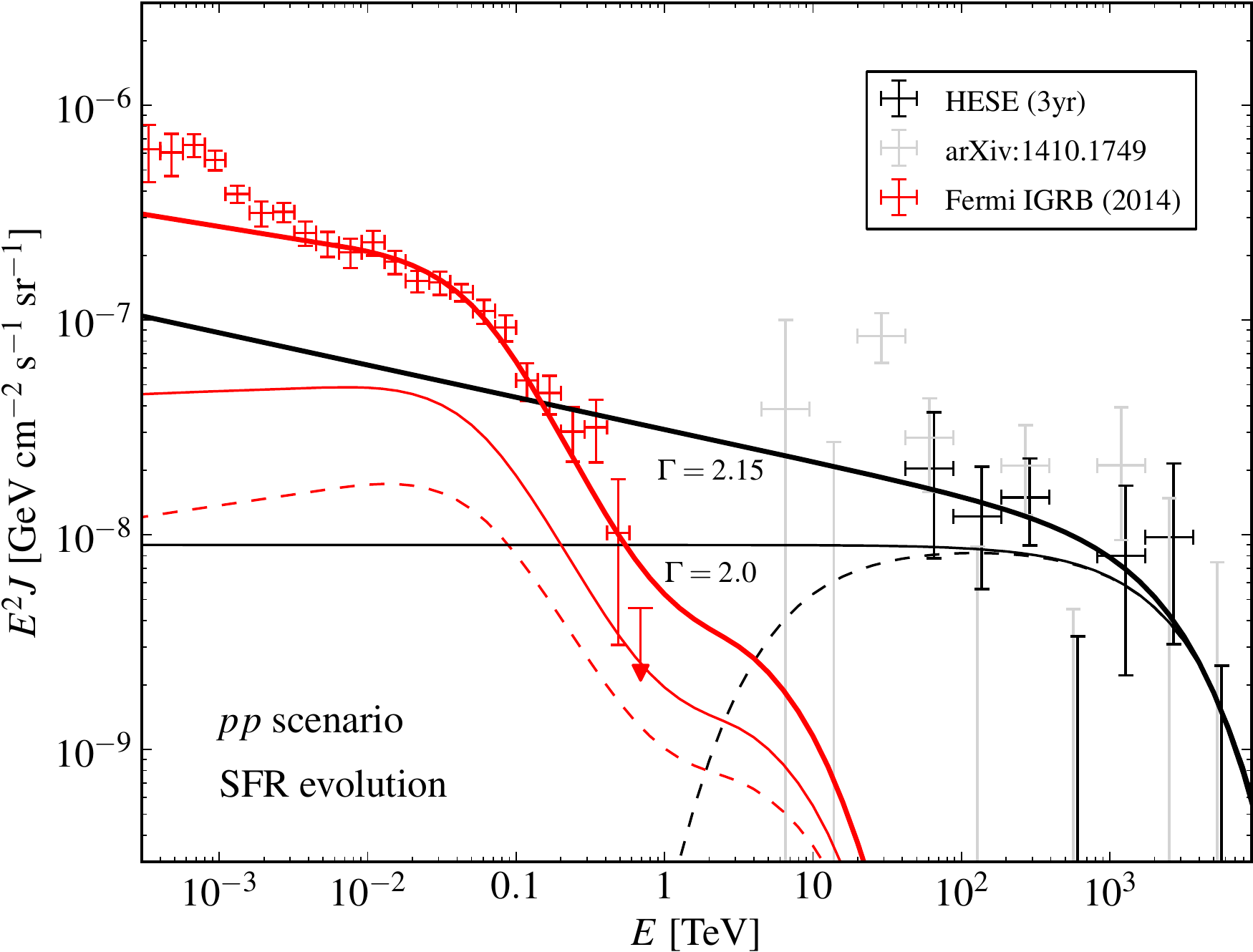}
\caption[]{Isotropic $\gamma$-ray background (IGRB) inferred by {\it Fermi}~\cite{Ackermann:2014usa} compared to the diffuse per-flavor neutrino flux observed by IceCube\cite{Aartsen:2013bka,Aartsen:2014muf} (updated plot of Ref.~\cite{Ahlers:2013xia}). The black lines show possible neutrino models consistent with the IceCube data. The red lines are the corresponding $\gamma$-rays of $pp$ scenarios reprocessed in the cosmic radiation background. The thick and thin solid lines show a power-law emission with $\Gamma=2.15$ and $\Gamma=2$, respectively, with an exponential cutoff around PeV. The dashed lines show an emission that is peaked in the 10TeV-PeV and only contributes in the $\gamma$-ray emission via cascades photons.} \label{fig3}
\end{figure}

Figure~\ref{fig3} shows three $pp$ emission scenarios that follow the diffuse neutrino observation in the TeV-PeV energy range. The black and red lines show the neutrino and $\gamma$-ray spectra after accounting for cosmic evolution and cascading in cosmic radiation backgrounds. The thick solid line shows the case of an emission following $E^{-2.15}$ with an exponential cutoff around PeV. This scenario is marginally consistent with the inferred isotropic diffuse $\gamma$-ray background (IGRB) by {\it Fermi}~\cite{Ackermann:2014usa}. The emission at sub-TeV energies is dominated by the direct photons of the sources. 

For harder emission ($\Gamma=2.0$, thin lines) the cascaded spectrum is still a significant contribution to the IGRB. The effect of cascades $\gamma$-rays is clearly visible as a bump in the GeV-TeV energy range. For illustration  we also show the effect of a low energy cutoff in the intrinsic $\gamma$-ray and neutrino spectra (dashed lines). As we already emphasized, this emission spectrum is not expected for a $pp$ scenario. However, the observed $\gamma$-ray spectrum is in this case dominated by secondary cascaded photons. The contribution to the {\it Fermi} IGRB between 100~GeV to 1~TeV is still at the level of 10\%. 

In general, this shows that the diffuse $\gamma$-ray contribution to the {\it Fermi} IGRB is large for $pp$ scenarios soft emission spectra ($\Gamma\gtrsim2.2$) are inconsistent with the data~\cite{Murase:2013rfa}. On the other hand, $p\gamma$ scenarios will most likely contribute to the {\it leptonic} emission of sources via reprocessed $\gamma$-rays. In this case, the hadronic counterparts of the IceCube observation can be identified in the source emission itself, but the energy range will depend on the particular source type.

\section{Galactic TeV-PeV $\gamma$-rays}

In the previous section we focused on the relation between CRs, $\gamma$-rays and neutrinos of extragalactic sources, which seem consistent with the absence of strong anisotropies in the observed neutrino spectrum. However, with the limited angular resolution and statistics of the observation it is possible that Galactic sources which are sufficiently extended contribute to the data. These extended emission regions are also observed by {\it Fermi} via the diffuse $\gamma$-ay emission of the Galactic Plane (GP)~\cite{FermiLAT:2012aa} or the extended {\it Fermi Bubbles} (FB)~\cite{Su:2010qj,Fermi-LAT:2014sfa}. In fact, as indicated in the sky map of Fig.~\ref{fig1} two of the PeV cascades (\#14 \& \#35) are within angular uncertainties consistent with an emission along the Galactic Plane and the weak cluster of cascades in an extended region around the Galactic Center might also indicate the presence of Galactic neutrino emission.

\begin{figure}[t]
\centering
\includegraphics[width=\linewidth]{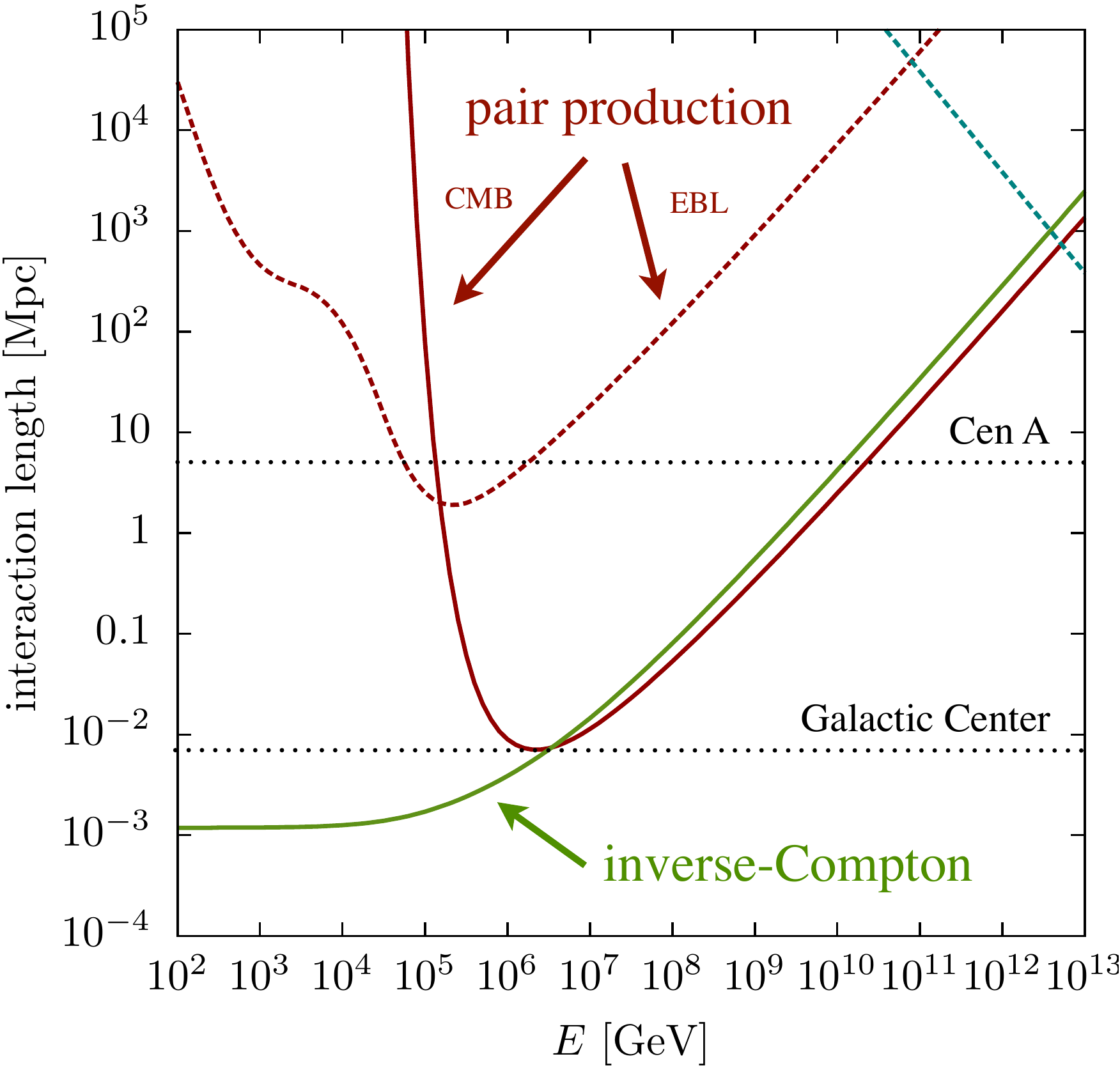}
\caption[]{The interaction length of pair production and inverse-Compton scattering of photons with the CMB and EBL. Typical distance scales like the Galactic Center and the close-by radio galaxy Cen A are indicated.} \label{fig4}
\end{figure}

Over Galactic distances the corresponding emission of hadronic TeV-PeV $\gamma$-rays are not completely attenuated by radiation backgrounds. In particular the observation of PeV $\gamma$-rays with an attenuation length of about 10 kpc via pair production in the cosmic microwave background (CMB) would be a {\it smoking gun} for Galactic production~\cite{Gupta:2013xfa,Ahlers:2013xia}. Figure~\ref{fig4} shows the interaction length of photons for pair production and inverse-Compton scattering of photons with the CMB and the extragalactic background light (EBL)~\cite{Franceschini:2008tp}. Extra-galactic candidate sources for PeV neutrino production, like Centaurus A at a distance of 4~Mpc shown in the plot, are only visible by hadronic $\gamma$-ray emission below 100~TeV. The diffuse flux of $\gamma$-rays from cosmic sources is only visible below 1~TeV due to EBL absorption.

\begin{figure}[t]
\centering
\includegraphics[width=\linewidth]{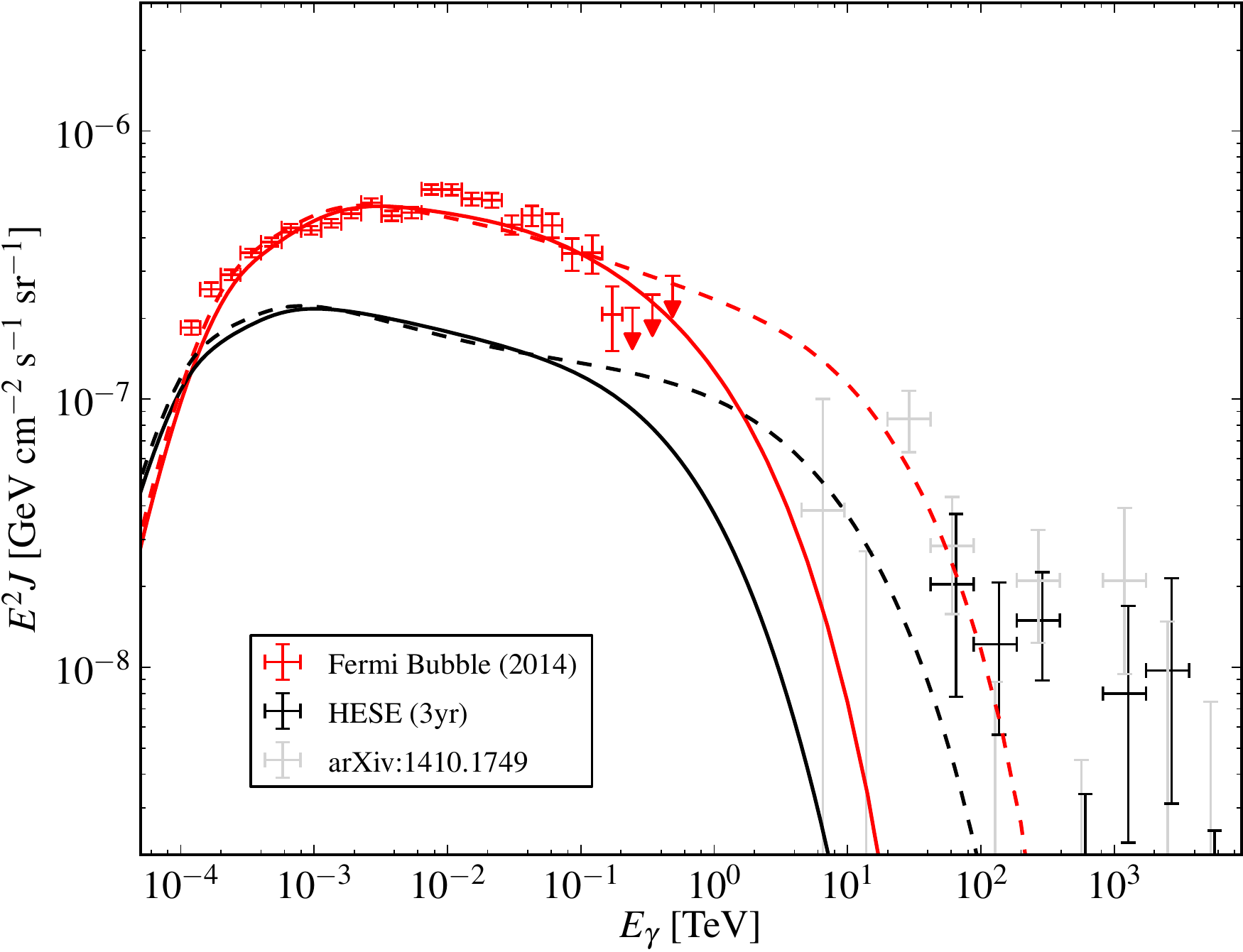}
\caption{The diffuse flux from the {\it Fermi Bubbles}~\cite{Fermi-LAT:2014sfa} compared to the diffuse per-flavor neutrino flux observed by IceCube\cite{Aartsen:2013bka,Aartsen:2014muf}. We show hadronic models of $\gamma$-ray (red lines) and per-flavor neutrino (black lines) emission. The lines show power-law emission of CR protons following the model in Ref.~\cite{Fermi-LAT:2014sfa} (Eq.~(16)) with $n=2.1$ and $E_{\rm cut}=13.7$~TeV (solid) or $n=2.15$ and $E_{\rm cut}=200$~TeV (dashed), respectively. In the case of a large cutoff the neutrino emission extends into the energy region studied in~\cite{Aartsen:2014muf}.} \label{fig5}
\end{figure}

The origin of the extended Galactic $\gamma$-ray emission known as the {\it Fermi Bubbles}~\cite{Su:2010qj} is unclear, but leptonic~\cite{Mertsch:2011es} as well as hadronic~\cite{Crocker:2010dg,Yang:2014pia} scenarios have been proposed, which can be distinguished via their corresponding neutrino emission~\cite{Lunardini:2011br}. Figure~\ref{fig5} shows the recent {\it Fermi} result of the emission spectrum of the FB region~\cite{Fermi-LAT:2014sfa}. The red lines shows possible hadronic emissions from a power-law CR spectrum with different spectral indices and exponential cutoffs assuming a $pp$ origin~\cite{Kamae:2006bf}. The black lines show the corresponding diffuse neutrino flux in comparison with the IceCube data. The models indicate that the extrapolated neutrino emission is probably irrelevant for PeV neutrino emission, but can have a noticeable contribution at energies of $(1-10)$~TeV~\cite{Aartsen:2014muf}. Note, that the extension of the {\it Fermi Bubbles} is only about 10\% of the full sky.

A guaranteed contribution to the diffuse emission of the Galactic Plane is the hadronic emission produced by interactions of diffuse CRs with gas~\cite{Ahlers:2013xia,Joshi:2013aua,Kachelriess:2014oma}. In general, this emission is expected to follow the local diffuse CR spectrum. Usually it is assumed that the average spectrum in our Galaxy is close to the observed one with a power-law $E^{-2.75}$ up to the {\it knee} at $(3-4)$ PeV where the spectrum softens. In this case the contribution to the diffuse neutrino flux at PeV is not expected to be significant. Nevertheless, some authors have argued that the average spectrum in our Milky Way might be harder and the locally observed spectrum might be softer due to a local and recent CR injection~\cite{Neronov:2014uma}. Again, this would not only produce an anisotropy of the neutrino emission along the GP, but also PeV $\gamma$-rays.

Exotic contributions like decaying heavy dark matter will also produce an extended emission~\cite{Feldstein:2013kka,Esmaili:2013gha,Bai:2013nga,Cherry:2014xra}. About 50\% of the Galactic signal will be within $60^\circ$ around the Galactic Center. It can be expected that these decaying dark matter scenarios leading to strong neutrino emission will also produce $\gamma$-rays up to an energy set by the mass scale. Interestingly, the neutrino emission of extragalactic dark matter decay will be at a similar flux level as the Galactic contribution. Hence, the high-energy neutrino events far off the Galactic Center can also be accounted for in this scenario without fine-tuning. 

\section{Summary and Outlook}

The first observation of high-energy astrophysical neutrinos have added an important new observable of multi-messenger astronomy. Their energy density is comparable to the power density of Galactic or extragalactic cosmic rays integrated over the Hubble timescale. It also similar to the energy density of the isotropic $\gamma$-ray background. These similarity might be the result of calorimetric processes and suggest that a large contribution of high-energy messengers have a hadronic origin.

The absence of strong anisotropies in the data can be a natural consequence of neutrino emission in extragalactic sources. The identification of individual sources via clusters in neutrino arrival directions is challenging due to the limited angular reconstruction, low signal statistics and large atmospheric backgrounds. Cross-correlation of neutrino events with catalogues of transient and continuous $\gamma$-rays sources will provide the best chance to 
identify the neutrino sources.

Interestingly, the isotropic diffuse $\gamma$-ray background observed by {\it Fermi}-LAT already constrains extragalactic hadronic emission scenarios. Neutrino production via cosmic ray interactions with gas ($pp$ scenario) predict neutrino and hadronic $\gamma$-ray spectra that follow the cosmic ray power-law spectrum. The tail of sub-TeV $\gamma$-rays for soft spectral indices $\Gamma\gtrsim2.2$ are inconsistent with the observed $\gamma$-ray background level. Harder emission scenarios can also be constrained by the identification of known diffuse $\gamma$-ray contributions, such as unresolved blazars.

Needless to say that neutrino astronomy would benefit from a larger instrument with an increased sensitivity for neutrino point sources. The proposed {\it IceCube--Gen2} extension~\cite{Aartsen:2014njl} plans to increase the effective volume of IceCube by about a factor of $10$. For transient sources which are not dominated by atmospheric backgrounds this would increase the sensitivity by about a factor of $10^{2/3}\simeq 5$.

\bigskip 
\begin{acknowledgments}
The author would like to thank the organizers of the {\it 5th Fermi Symposium} 2014 for a very pleasant and stimulating conference. This work is supported by the National Science Foundation under grants OPP-0236449 and PHY-0236449.
\end{acknowledgments}

\bigskip 
\bibliographystyle{h-physrev}
\bibliography{bib}





\end{document}